\documentclass{natureJLH}
\usepackage[english]{babel}
\usepackage{blindtext}
\usepackage{epsfig}
\usepackage{subcaption} 
\usepackage{ae, aecompl}
\usepackage{helvet}
\usepackage{epsfig}
\usepackage{dcolumn}
\usepackage{color}
\usepackage[LGR, T1]{fontenc}
\usepackage[latin9]{inputenc}
\usepackage{amssymb}
\usepackage{amsmath}
\usepackage{mwe, tikz}
\usepackage{esint}
\usepackage{longtable}
\usepackage{url}
\usepackage{etoolbox}
\usepackage{multirow}

\newcommand\apj{Astrophys. J.,}
\newcommand\apjl{Astrophys. J. Lett.,}

\newcommand\mnras{Mon. Not. R. Astron. Soc.,}
\newcommand\nat{Nature,}

\usepackage{xpatch}

\title{Astrophysics: Extreme emission see from $\gamma$-ray bursts} 

\author{
Bing Zhang
}

\begin{document}

\maketitle

\begin{affiliations}
\item Department of Physics and Astronomy, University of Nevada, Las Vegas, NV 89154, USA. Email: zhang@physics.unlv.edu
\end{affiliations}


\begin{abstract}
Cosmic explosions called $\gamma$-ray bursts are the most energetic bursting events in the universe. Observations of extremely high-energy emission from two $\gamma$-ray bursts provide a new way to study these gigantic explosions.
\end{abstract}

Astrophysical explosions known as $\gamma$-ray bursts (GRBs) can release in one second the amount of energy that the Sun will produce in its enture lifetime\cite{zhang18}. The emission from GRBs covers a wide range of the electromagnetic spectrum and occurs in two stages: the prompt-emission phase and the afterglow phase. The main emission mechanism is thought to be synchrotron radiation, whereby the gyration of energetic electrons around magnetic-field lines releases photons. Until now, emission from GRBs has been observed only at energies below 100 gigaelectronvolts (GeV). Three papers in this issue\cite{MAGIC1,MAGIC2,HESS} report observations of $\gamma$-rays that have energies above 100 GeV from two bright GRBs, dubbed GRB 190114C and GRB 180720B. 

The Major Atmospheric Gamma Imaging Cherenkov (MAGIC) Collaboration\cite{MAGIC1} (page 455) detected photons in the teraelectronvolt range (1 TeV is $10^3$ GeV) from GRB 190114C, using the MAGIC telescopes at La Palma, Spain. The first detections started about one minute after the burst triggered NASA's two space-borne GRB detectors: the Burst Alert Telescope onboard the Swift satellite and the Gamma-ray Burst Monitor onboard the Fermi satellite. The high-energy photons continued to rain down on the MAGIC telescopes for about 20 minutes, with the flux decreasing rapidly over time. The MAGIC Collaboration and colleagues\cite{MAGIC2} (page 459) detected this GRB using several other ground-based and space-borne telescopes. When combined with the MAGIC data, this rich data set allowed the authors to comprehensively model  the event and study how the TeV emission was produced.

Abdalla et al.\cite{HESS} (page 464) detected photons of energies above 100 GeV (but below 1 TeV) from GRB 180720B, using the High Energy Stereoscopic System (HESS) array of telescopes in Namibia. Although these photons were lower in energy and fewer in number than those observed from GRB 190114C, they were detected from deep in the afterglow phase (10 hours after the GRB was triggered and lasting for 2 hours).  The flux and maximum energy of the afterglow emission decrease over time, owing to deceleration of the jets -- the two narrow, oppositely directed channels through which most of the explosive energy of a GRB is released. Consequently, the detection of such high-energy photons deep in the afterglow phase is also groundbreaking. 

The MAGIC and HESS observatories both use an array of optical telescopes called imaging atmospheric Cherenkov telescopes (IACTs), which are designed to detect $\gamma$-rays in the very-high-energy range (roughly from 30 GeV to 100 TeV). More precisely, the IACTs detect the light (known as Cherenkov radiation) that is produced when such $\gamma$-rays hit Earth's atmosphere and produce a shower of charged particles. These facilities have been operating for more than a decade. GRBs, as the most powerful explosions in the universe, have been one of the main observational targets, but, until now, have evaded detection. The current results are therefore a triumph for these observatories.

The discoveries are also a triumph for GRB theories. Theoretically, there are three mechanisms by which high-energy $\gamma$-rays can be produced during the afterglow phase\cite{zhang01}. The first is synchrotron radiation from electrons accelerated by the external shock -- the shock wave that is generated when the exploded matter collides with surrounding interstellar gas. This emission component has a maximum energy that depends only on the Lorentz factor of the outflow (a parameter that denotes how fast the external shock is moving). To reach energies above $100$ GeV, the Lorentz factor must be greater than\cite{zhang01} about $1,000$, which is only marginally possible.  Observations show that the Lorentz factor of GRB jets is usually a few hundred during the prompt-emission phase and decreases over time during the afterglow phase\cite{racusin11}. 

The second high-energy radiation mechanism is synchrotron radiation from protons accelerated by the GRB external shock. This emission component can, in principle, contain TeV $\gamma$-rays. However, because protons are much less efficient emitters than are electrons, the conditions for this component to be dominant are rather demanding. Finally, the third mechanism is called synchrotron self-Compton\cite{sari01} (SSC), whereby the same accelerated electrons that emit synchrotron photons can scatter off some of these photons, resulting in photons that have energies above $100$ GeV (Fig.1a). For typical shock-microphysics parameters inferred from afterglow modeling of other GRBs, it is expected that the SSC component should usually be the main way in which high-energy $\gamma$-rays are produced\cite{zhang01}.

One key prediction of the SSC mechanism is that there should be two `humps' in the spectral energy distribution (SED) of the afterglow spectrum\cite{sari01,zhang01} (Fig.1b). Such a two-hump structure has been commonly observed for high-energy jets launched from supermassive black holes known as blazars\cite{ghisellini17}, and the same structure has been confidently expected for GRBs. Previous observations of high-energy afterglows of GRBs using the Large Area Telescope onboard the Fermi satellite have not convincingly shown the existence of a second hump in the spectral energy distributions\cite{kumar09}. However, some tentative evidence has been collected from another bright burst\cite{ackermann14,liu13,fan13}, GRB 130427A.

The MAGIC Collaboration and colleagues' multi-wavelength observations of GRB 190114C have  firmly established, for the first time, the existence of the SSC component in a GRB afterglow\cite{MAGIC2}. This conclusion has been confirmed by independent modeling from other groups\cite{wang19,derishev19,fraija19}. The double-hump feature is comparatively less clear in the spectral energy distribution obtained by Abdalla et al. for GRB 180720B. However, in the late afterglow phase, electron synchrotron radiation cannot produce photons of energies above $100$ GeV without the need to introduce exotic particle-acceleration mechanisms. As a result, the SSC mechanism is the preferred explanation for the observed spectral energy distribution\cite{HESS,wang19}. 

Why did it take so long to detect a theoretically expected common spectral component? The observation of a GRB by an IACT requires that the burst is bright (to produce a sufficient number of high-energy photons) and nearby (to avoid absorption of the photons by the infrared background radiation in the Universe). Furthermore, the telescope needs to have the correct observational conditions. For instance, a particular GRB would not be detected by an IACT if the event occurred during the daytime, in poor weather or in the area of the sky that was not accessible by the telescope.  Nevertheless, the breakthrough results reported in the current papers suggest that, with dedication and probably a bit of luck, a revolutionary discovery can be made. 

Now that photons of energies above $100$ GeV have been detected from GRBs, it is expected that such detections will become routine in the future -- especially with the full operations of the available IACTs and of observatories that use other detection techniques, such as the High-Altitude Water Cherenkov Observatory in Maxico. The field will also greatly benefit from  the operations of facilities such as the Cherenkov Telescope Array and the Large High Altitude Air Shower Observatory in Daocheng, China. As history has repeatedly shown, the opening of a new spectral window in GRB research always reveals many treasures for researchers to mine. This spectral window at the highest energies will not be any different, and can be even more rewarding. 





\newpage

\begin{figure}[t]
\begin{tabular}{c}
\includegraphics[keepaspectratio, clip, width=0.75\textwidth,angle=-90]{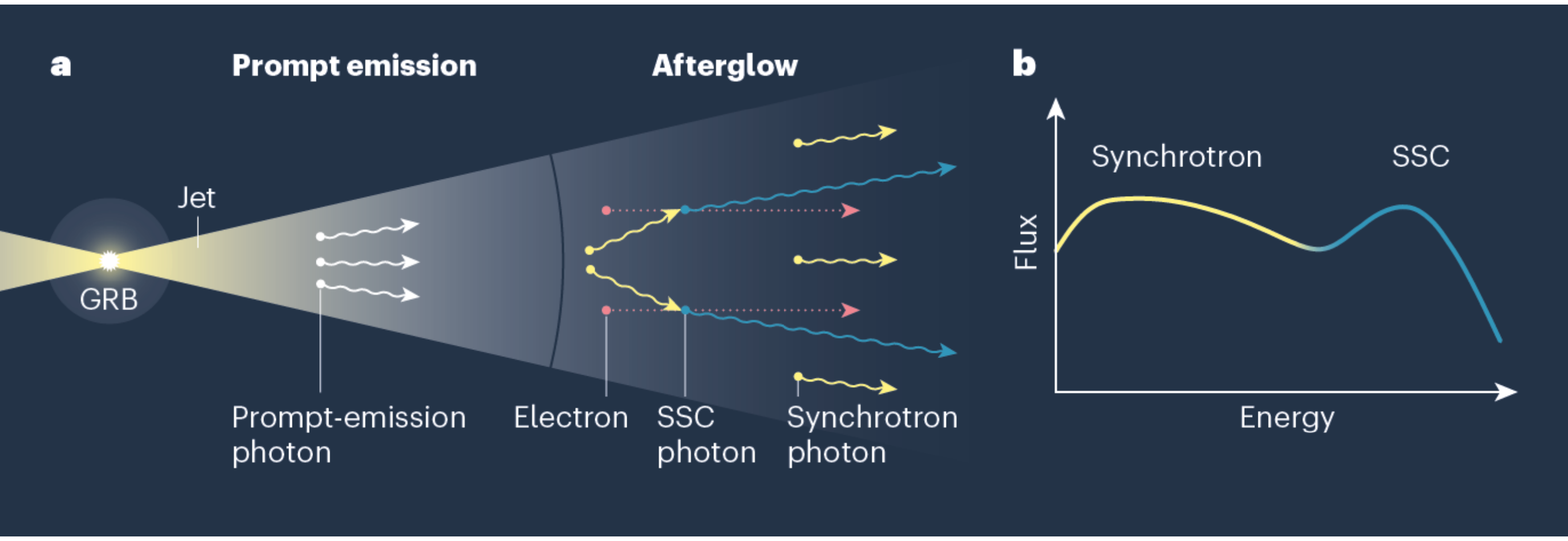} \\
\end{tabular}
\caption{{\bf  Emission from a $\gamma$-ray burst}. {\bf a.} Three papers\cite{MAGIC1,MAGIC2,HESS} report the detection of high-energy radiation from astrophysical explosions known as $\gamma$-ray bursts (GRBs). The explosive energy from a GRB is thought to be channelled into two narrow jets. The emission of photons occurs in two stages: the prompt-emission phase and the afterglow phase. In the afterglow phase, photons can be generated through a mechanism called synchrotron radiation. High-energy photons are thought to be mainly produced through a process dubbed synchrotron self-Compton (SSC), whereby the scattering of synchrotron photons off energetic electrons gives the photons a boost in energy\cite{zhang01,sari01}. {\bf b.} One key prediction of the SSC mechanism is that there should be two humps in the spectral energy distribution of the afterglow spectrum: one corresponding to synchrotron photons and the other to SSC photons\cite{zhang01,sari01}. Results from the three papers firmly establish the existence of such an SSC component.}
\label{fig:lc_sp}
\end{figure}

\end{document}